\documentclass[12pt]{article}
\usepackage{axodraw}
\usepackage[dvips]{graphicx}
\def\NPB#1#2#3{{\em Nucl. Phys.} {\bf B#1} (19#2) #3}
\def\PLB#1#2#3{{\em Phys. Lett.} {\bf B#1} (19#2) #3}

\def\PRT#1#2#3{{\em Phys. Rep.} {\bf#1} (19#2) #3}

\def\ZPC#1#2#3{{\em Zeit. f\"ur Physik} {\bf C#1} (19#2) #3}

\begin{document}
\title{Associate Higgs and Gauge Boson Production  at Hadron Colliders
  in a Model with Vector Resonances} 

\author{ Alfonso R. Zerwekh
  \footnote{alfonsozerwekh@uach.cl} \\
 Instituto de F{\'{\i}}sica, Facultad de Ciencias \\
Universidad Austral de Chile \\ Casilla 567, Valdivia, Chile}
\date{}

\maketitle

\begin{abstract}
  Motivated by new models of dynamical electroweak symmetry breaking
  that predict a light composite higgs boson, we build an effective
  lagrangian which describes the Standard Model (with a light Higgs)
  and vector resonances. We compute the cross section for the associate
  production of a higgs and a gauge boson. For some values of  model
  parameters we find that the cross section is significantly enhanced with
  respect to the Standard Model. This enhancement  is similar at the
  LHC and the Tevatron for the same range of resonance mass.   
\end{abstract}

\section{Introduction}
\label{sec:intro}

One of the main challenges faced today by Particle Physics is to
elucidate the nature of electroweak symmetry breaking. This is the
only aspect of the Standard Model that still  has not been directly
tested by experiment. Moreover, there is a general agreement in the
sense that, due to the hierarchy and triviality problems, the standard
higgs sector is not satisfactory and it  really points to
the existence of new physics at the TeV scale.   

One possibility for such new physics is the existence of a new
strong interaction that dynamically breaks the electroweak symmetry
\cite{technicolor}. The
best known realization of this elegant idea, technicolor, has 
been challenged by LEP precision measurements. In fact, the original
models, which are scaled up versions of QCD, are ruled out. A great
progress toward the construction of potentially more successful models was
the introduction of walking technicolor which basically means that the
interaction is quasi-conformal over a large range of
energy. Nevertheless, in general these models predict that the composite
scalar (the higgs) in the spectrum must be heavy, while the
current experimental data seem to point to the existence of a light
higgs.

Very recently a new kind of technicolor models has been proposed
\cite{newmodels} whose
main characteristic is that technifermions are not in the fundamental
representation of the technicolor group. In these models the walking
behavior of the coupling constant appears naturally and they are not in
conflict with the current limits on the oblique parameters. But the
most remarkable feature of these models is that they predict the
existence of light composite higgs with a mass around 150 GeV.

Inspired by such models, we write down an effective lagrangian which
describes the Standard Model with a light higgs and vector resonances
which are a general prediction of dynamical symmetry breaking models
\cite{dominici}. 
The model is minimal in the sense that we assume that any other
composite state would be heavier than the vector resonances, and so
they are not taken into account, and there are no physical technipions in the
spectrum.

 In what follows we will study some aspects of such a model. Section
 \ref{sec:lag} 
 is devoted to the construction of the lagrangian. In section \ref{sec:res} we
 compute the cross section for the associate production of a higgs
 and a gauge boson and we compare the results with the Standard Model.
 Finally in section \ref{sec:con} we present our conclusions.

\section{The Lagrangian}
\label{sec:lag}

\subsection{Gauge Sector}
\label{sec:gaugesector}

We start by noticing that, in general, dynamical electroweak symmetry
breaking models predict the existence of composite vector particles (
the so called technirho and techniomega) that mix with the gauge
bosons of the Standard Model. In order to describe this mixing, we use
a generalization of  Vector Meson Dominance \cite{VMD} introduced
in  \cite{zerwekh_and_rosenfeld} and developed 
in \cite{zerwekh}. In this approach we choose a representation where
all vector fields transform as gauge fields and they mix through a mass
matrix. On the other hand, gauge invariance imposes that the mass
matrix has a null determinant. In our case, the lagrangian for the
gauge sector can be written as:

\begin{eqnarray}
  \label{eq:vectlag}
  {\cal L}&=& -\frac{1}{4}W^{a}_{\mu \nu}W^{a \mu \nu}
  -\frac{1}{4}\tilde{\rho}^{a}_{\mu \nu}\tilde{\rho}^{a \mu \nu}
  \nonumber \\
 & & + \frac{M^2}{2}\left(
  \frac{g^2}{g^2_2}W^a_{\mu}W^{a\mu}+\tilde{\rho}^{a}_{\mu}\tilde{\rho}^{a\mu}-
\frac{2g}{g_2}W^a_{\mu}\tilde{\rho}^{a \mu}\right) \nonumber \\
  & &-\frac{1}{4}B_{\mu \nu}B^{\mu \nu}
  -\frac{1}{4}\tilde{\omega}_{\mu \nu}\tilde{\omega}^{\mu \nu}
  \nonumber \\
& & + \frac{M'^2}{2}\left(
  \frac{g'^2}{g'^2_2}B_{\mu}B^{\mu}+\tilde{\omega}_{\mu}\tilde{\omega}^{\mu}-
\frac{2g'}{g'_2}B_{\mu}\tilde{\omega}^{\mu}\right) 
\end{eqnarray}
where

\begin{eqnarray}
  \label{eq:def1}
  W^{a}_{\mu
  \nu}&=&\partial_{\mu}W^a_{\nu}-\partial_{\nu}W^a_{\mu}+g\epsilon^{abc}
  W^b_{\mu} W^c_{\nu} \\
\tilde{\rho}^{a}_{\mu
  \nu}&=&\partial_{\mu}\tilde{\rho}^a_{\nu}-\partial_{\nu}\tilde{\rho}^a_{\mu}
+g_2\epsilon^{abc}
  \tilde{\rho}^b_{\mu} \tilde{\rho}^c_{\nu} \\
 B_{\mu
  \nu}&=&\partial_{\mu}B_{\nu}-\partial_{\nu}B_{\mu}\\
\tilde{\omega}_{\mu
  \nu}&=&\partial_{\mu}\tilde{\omega}_{\nu}-\partial_{\nu}\tilde{\omega}_{\mu}
\end{eqnarray}
and $M$ ($M'$) is the hard mass of the proto-technirho
(proto-techniomega). Notice that our lagrangian is written in term of
non-physical fields. The physical ones will be obtained by
diagonalizing the mass matrix.

By construction, lagrangian (\ref{eq:vectlag}) is invariant under
$SU(2)_{\mbox{L}} \times U(1)_{\mbox{Y}}$. The symmetry breaking to
$U(1)_{\mbox{em}}$ will be described by mean of the vacuum expectation
value of a scalar field, as in the Standard Model. In other words, we
will use an effective gauged linear sigma model as a phenomenological
description of the electroweak symmetry breaking.      

\subsection{Fermions}
\label{sec:fermions}

As usual, fermions are minimally coupled to gauge  bosons through a
covariant derivative. Because in our scheme all the vector bosons
transform as gauge fields, it is possible to include the
proto-technirho and the proto-techniomega in an ``extended'' covariant
derivative \cite{zerwekh}, resulting in the following lagrangian for
the fermion sector:

\begin{equation}
  \label{eq:fermionlag}
  {\cal
  L}=\bar{\psi_L}i\gamma^{\mu}D_{\mu}\psi_L+\bar{\psi_R}i\gamma^{\mu}
\tilde{D}_{\mu}\psi_R   
\end{equation}
with
\begin{eqnarray}
  D_{\mu}&=&\partial_{\mu}+ i\tau^a g(1-x_1)W^a_{\mu}+ 
i\tau^a g_2 x_1 \tilde{\rho}^a_{\mu} \nonumber \\
& &+ i\frac{Y}{2} g'(1-x_2)B_{\mu} + i\frac{Y}{2} g'_2 x_2
\tilde{\omega}_{\mu}  
\end{eqnarray}
and
\begin{equation}
 \tilde{D}_{\mu}=\partial_{\mu}+ i\frac{Y}{2} g'(1-x_3)B_{\mu} +
 i\frac{Y}{2} g'_2 x_3 
\tilde{\omega}_{\mu}  
\end{equation}

The parameters $x_i$ ($i=1,2,3$) play a role similar to fermion
delocalization in deconstruction models \cite{Chivukula:2005bn}.

A direct coupling between fermions and the vector resonances can
appear naturally in technicolor due to extended technicolor
interactions. Nevertheless, they must be proportional to the fermion
mass and then,they are not important except for the top quark. 
In what follows, for simplicity, we take $x_1=x_2=x_3=0$.

 \subsection{Higgs Sector and Mass Matrices}
\label{sec:higgs}

In our effective model the higgs sector is assumed to be the same as
in 
the Standard Model except by the possibility of including a direct
coupling between the higgs doublet and the vector resonances through an
``extended'' covariant derivative as was shown for the fermion
sector. That is, the lagrangian for the Higgs sector is:

\begin{equation}
  \label{eq:Higgs}
  {\cal L}= \left(D^{\mu}\Phi \right)^{\dagger}\left(D_{\mu}\Phi
  \right)-V\left(\Phi \right) 
\end{equation}
where, as usual

\begin{equation}
  \label{eq:potential}
  V\left(\Phi \right)=-\mu^2 \Phi^{\dagger}\Phi+\lambda\left(
  \Phi^{\dagger}\Phi \right)^2 
\end{equation}
and

\begin{eqnarray}
  D_{\mu}&=&\partial_{\mu}+ i\tau^a g(1-f_1)W^a_{\mu}+ 
i\tau^a g_2 f_1 \tilde{\rho}^a_{\mu} \nonumber \\
& &+ i\frac{Y}{2} g'(1-f_2)B_{\mu} + i\frac{Y}{2} g'_2 f_2
\tilde{\omega}_{\mu}  
\end{eqnarray}

This direct coupling may seem natural in a context where the higgs and
the vector resonances are both composite states of the same underlying
strong sector. Nevertheless, in this work we will avoid it (we choose
$f_1=f_2=0$) because, in principle, it can introduce dangerous tree
level corrections to the $\rho$ parameter.

Once the electroweak symmetry has been broken, the mass matrix of the vector
bosons takes contributions from (\ref{eq:vectlag}) and from the Higgs
mechanism. For the neutral vector bosons, the resulting mass matrix is:

\begin{equation}
  \label{eq:neutralmassmatrix}
  {\cal M_{\mbox{neutral}}}=\frac{v^2}{4}\left [
    \begin{array}[h]{c c c c}
(1+\alpha)g^2 & -\alpha g g_2 & -g g' & 0 \\
-\alpha g g_2 & \alpha g_2^2 & 0 & 0 \\
-g g' & 0 & (1+\alpha ')g'^2 & - \alpha ' g' g_2'\\
0 & 0 &  - \alpha ' g' g_2' & \alpha ' g'^2 
    \end{array}
\right ]
\end{equation}
where

\begin{equation}
  \label{eq:alpha}
  \alpha=\frac{4M^2}{v^2 g^2_2}
\end{equation}
and
\begin{equation}
  \label{eq:alphaprima}
  \alpha '=\frac{4M'^2}{v^2 g'^2_2}
\end{equation}

In what follows, in order to simplify our analysis, we will assume that
$\alpha ' = \alpha $ and $ g_2' = g_2 $. Notice that this assumption
implies that the technirho and the techniomega will have the same
mass. We will also assume that $M$ 
is proportional to $g_2^2$ which is quit natural when we realize that
$M$ is a dynamical mass produced by the underlying strong interaction.

On the other hand, the mass matrix for the charged vector bosons can
be written as:

\begin{equation}
  \label{eq:chargedmassmatrix}
  {\cal M_{\mbox{charged}}}=\frac{v^2}{4}\left [
    \begin{array}[h]{c c}
(1+\alpha)g^2 & -\alpha g g_2 \\
-\alpha g g_2 & \alpha g_2^2 
    \end{array}
\right ]
\end{equation} 

We diagonalize the mass matrices in the limit $g_2 \rightarrow \infty$
and we obtain the following expressions for the physical fields (we
have dropped the Lorentz indexes).
We write the transformation to physical fields in the limit $g/g_2 \ll
1$ and we keep terms up to order $g/g_2$.

\begin{eqnarray*}
  \label{eq:eigenvectors}
  A&=&\frac{g'}{\sqrt{g^2+g'^2}}W^3+\frac{g}{\sqrt{g^2+g'^2}}B+\frac{g
  g'}{g_2\sqrt{g^2+g'^2}}\tilde{\rho}^3+ \frac{g
  g'}{g_2\sqrt{g^2+g'^2}}\tilde{\omega}\\
  Z&=&\frac{g}{\sqrt{g^2+g'^2}}W^3-\frac{g'}{\sqrt{g^2+g'^2}}B+\frac{g^2
  }{g_2\sqrt{g^2+g'^2}}\tilde{\rho}^3- \frac{
  g'^2}{g_2\sqrt{g^2+g'^2}}\tilde{\omega}\\
  \rho^0&=&-\frac{g}{g_2}W^3+\tilde{\rho}^3\\
  \omega&=&-\frac{g'}{g_2}B+\tilde{\omega}\\
   W^{\pm}&=&\tilde{W}^{\pm}+\frac{g}{g_2}\tilde{\rho}^{\pm}\\
   \rho^{\pm}&=&\tilde{\rho}^{\pm}-\frac{g}{g_2}\tilde{W}^{\pm}
\end{eqnarray*}
where
$$
\tilde{W}^{\pm}=\frac{1}{\sqrt{2}}\left(W^1 \mp i W^2\right)
$$
and
$$
\tilde{\rho}^{\pm}=\frac{1}{\sqrt{2}}\left(\rho^1 \mp i \rho^2\right)
$$

\begin{figure}[htbp]
  \centering
  \includegraphics[scale=1.5]{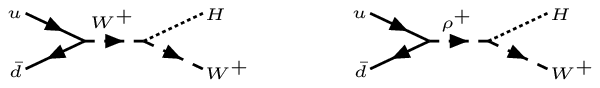}\\
  \begin{picture}(2,2)
    \Text(1,1)[b]{a)}
  \end{picture}\\

  \includegraphics[scale=1.5]{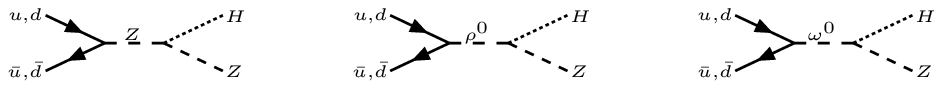}\\
  \begin{picture}(2,2)
    \Text(1,1)[b]{b)}
  \end{picture}
  \caption{Feynman diagrams for the production of a Higgs and a
    $W^+$ (a) , and a  Higgs and a $Z$ (b)}
  \label{fig:diagrams}
\end{figure}

The relevant Feynman rules, in terms of the physical
fields, for the associate production of a higgs and a gauge boson can be found
in table \ref{tab:FeynmanRules} while figure \ref{fig:diagrams} shows
the Feynman diagrams for the processes studied in this work.

\begin{table}[htbp]
  \centering
  \begin{tabular}{l|l}\hline
Fields in the vertex & Variational derivative of Lagrangian by fields
\\ \hline 
 $H$ $\omega^0_{\mu }$ ${Z}_{\nu }$&
        $\frac{1}{2}\frac{ e{}^2 M_W \sqrt{\alpha} v}{ c_w{}^3 M_{\rho}}g^{\mu
          \nu}$\\
$H$ $\rho^0_{\mu }$ ${Z}_{\nu }$ &
        $-\frac{1}{2}\frac{ e{}^2  M_W \sqrt{\alpha} v}{ c_w{}^2
          M_{\rho} s_w}g^{\mu
          \nu} $\\ 
$H$ $\rho^+_{\mu }$ $W^-{}_{\nu }$  &
        $-\frac{1}{2}\frac{ e{}^2 M_W \sqrt{\alpha} v}{ M_{\rho}
          s_w{}^2 } g^{\mu \nu} $\\
$\bar{u}$ $d$ $\rho^+_{\mu }$&
        $\frac{1}{8}\frac{ e{}^2 \sqrt{2} \sqrt{\alpha}  Vud v}{
          M_{\rho} s_w{}^2}(1-\gamma^5)  \gamma^\mu $\\  
$\bar{u}$ $u$ $\omega^0_{\mu }$ &
        $\frac{1}{24}\frac{ e{}^2 \sqrt{\alpha} v}{ c_w{}^2 M_{\rho}}
        \gamma^\mu\big((1-\gamma^5) +4(1+\gamma^5) \big)$\\  
$\bar{u}$ $u{}$ $\rho^0_{\mu }$ &
        $\frac{1}{8}\frac{ e{}^2 \sqrt{\alpha} v}{ c_w  M_{\rho}  s_w}
        (1-\gamma^5)\gamma^\mu $\\ 
\hline 
  \end{tabular}
  \caption{Feynman Rules for the relevant couplings of the vector resonances
    for the associated production of a higgs and a gauge
    bosons. The couplings of the $W^{\pm}$ and $Z$ to the quarks  are
    identical, in our limit, to the SM}
  \label{tab:FeynmanRules}
\end{table}

We have build our lagrangian based on an extension of vector meson
dominance. Another possibility for studying models with vector
resonances is to consider the so called  hidden local symmetry\cite{bando}. In
this case, the model would correspond to a nonlinear sigma model
based on the coset space $U(2)_L\otimes
U(2)_R/U(2)_V$. The
archetype of this kind of model applied to a strong electroweak model
is the so called BESS model \cite{dominici,mbess} (which is based on $(SU(2)_L
\otimes SU(2)_R)/SU(2)_V$ and hence does not have a techniomega). Of
course such a model does not include a higgs, nevertheless a
``linear'' version of the BESS model exists \cite{lbess} which include
scalars in the spectrum.       

\section{Results}
\label{sec:res}

We use LanHEP \cite{LanHEP} and CompHEP \cite{CompHEP} in order to
compute the cross section of the associate production of a higgs and a
gauge boson at the LHC and the Tevatron (Run II). We choose to work
with $\alpha=0.1$ because for values of $\alpha$ of this order, the
vector resonances can be light (i.e. $M_{\rho} \approx 250$ GeV) while
$g_2/g$ is still much bigger than one. As $\alpha$ approaches to one,
the vector resonances became too heavy and their observation is
increasingly difficult. On the other hand, if $\alpha$ is too small
the coupling of the vector resonances to the SM fields are suppressed.

The decay width of the vector resonances were also computed with CompHEP.   
In the range of masses considered here, the techniomega and the charged
technirho decay 
mainly into fermions
 while the neutral technirho decays mainly into a 
pair of $W$'s.

Notice that $\rho W Z$ and $\rho W A$ vertices do not exist in this 
model. Indeed, gauge invariance avoid the generation of such couplings
from Yang-Mills kinetic terms. For example, a $\rho W A$ vertex
coming from Yang-Mills term would violate gauge symmetry in the
process $\gamma W \rightarrow \gamma W$. A discussion about this
coupling in a similar model can be found in \cite{rosenfeld}.

Our aim in this work is to study the impact of the presence of vector
resonances on the higgs production in association with a gauge
boson. In particular, we want to compare the predictions of this
scenario with the Standard Model. For this reason we define the
following quantity as a measure of the enhancement of the signal over
the Standard Model: 

\begin{equation}
  \label{eq:r}
  \epsilon=\frac{\sigma-\sigma_{\mbox{SM}}}{\sigma_{\mbox{SM}}}
\end{equation}
\noindent
where $\sigma$ is the cross section predicted by
our effective lagrangian and $\sigma_{\mbox{SM}}$ is the cross section
predicted by the Standard Model. 

\begin{figure}[htbp]
  \centering
  \includegraphics{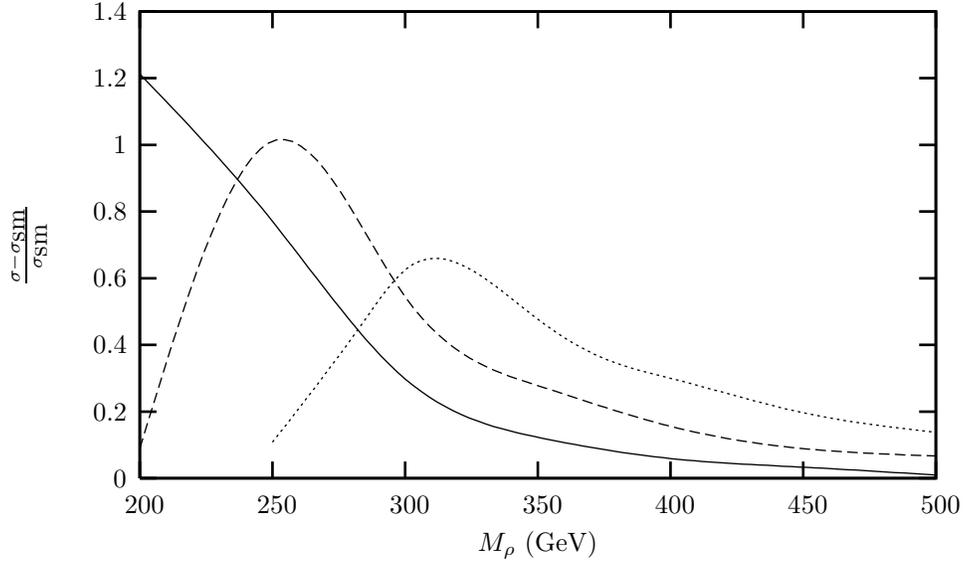}
  \caption{Enhancement of the cross section in the process $pp
  \rightarrow W^+ H$ at the LHC for three values of the higgs mass:
  $M_H=115$ GeV (solid line), $150$ GeV (dashed line) and $200$ GeV
  (pointed line)}
  \label{fig:lhc_wh}
\end{figure}

\begin{figure}[htbp]
  \centering
  \includegraphics{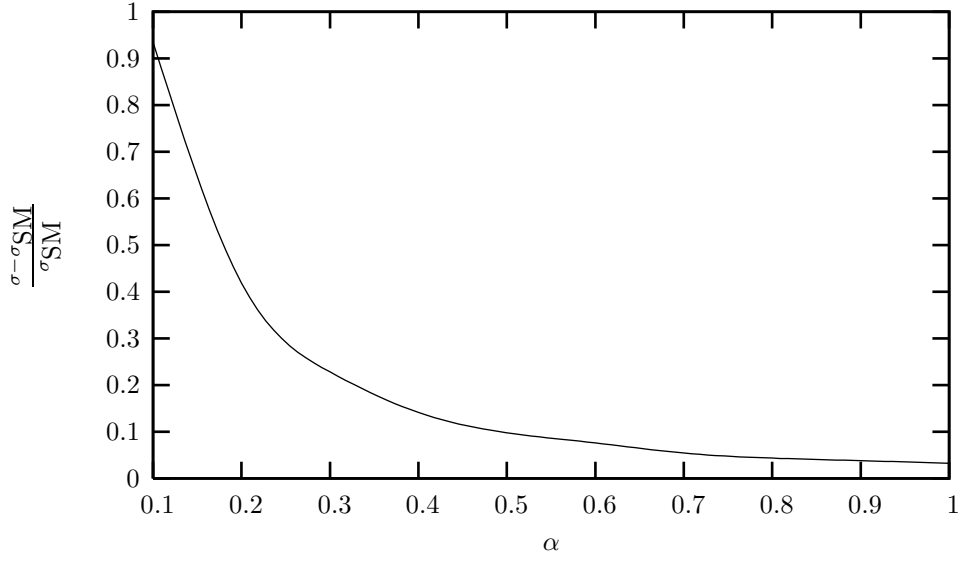}
  \caption{Enhancement of the cross section in the process $p\bar{p}
  \rightarrow W^+ H$ at the LHC as a function of $\alpha$ for
  $M_H=150$ GeV and $g/g2=0.1$ }
  \label{fig:alfa}
\end{figure}

In figure \ref{fig:lhc_wh} we show the value of $\epsilon$ as a function of
the mass of the technirho ($M_{\rho}$) for three values of the higgs
mass ($M_H=115$ GeV (solid line),$150$ GeV (dashed line) and $200$ GeV
(dotted line)) for the process $pp \rightarrow H W^+$ at the
LHC. Observe that in this case, the cross section is
significatively enhanced with respect to the standard model when the
technirho has a mass between 200 GeV and 350 GeV. The variation of
this enhancement as function of $\alpha$ is shown in figure
\ref{fig:alfa} for $M_H=150$ GeV and $g2/g=10$ 

\begin{figure}[htbp]
  \centering
  \includegraphics{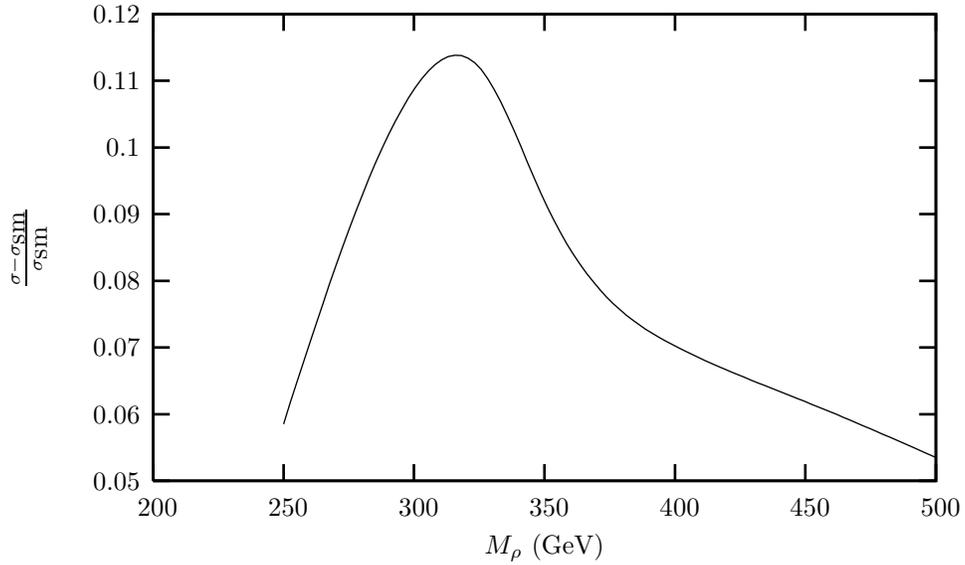}
  \caption{Enhancement of the cross section in the process $pp
  \rightarrow Z H$ at the LHC for $M_H=200$ GeV}
  \label{fig:lhc_zh}
\end{figure}

 On the other hand,
when a higgs and a $Z$ are produced (figure \ref{fig:lhc_zh}), the
cross section is less enhanced and we expect that this channel will
not be sensible to the presence of the vector resonances.   

\begin{figure}[htbp]
  \centering
  \includegraphics{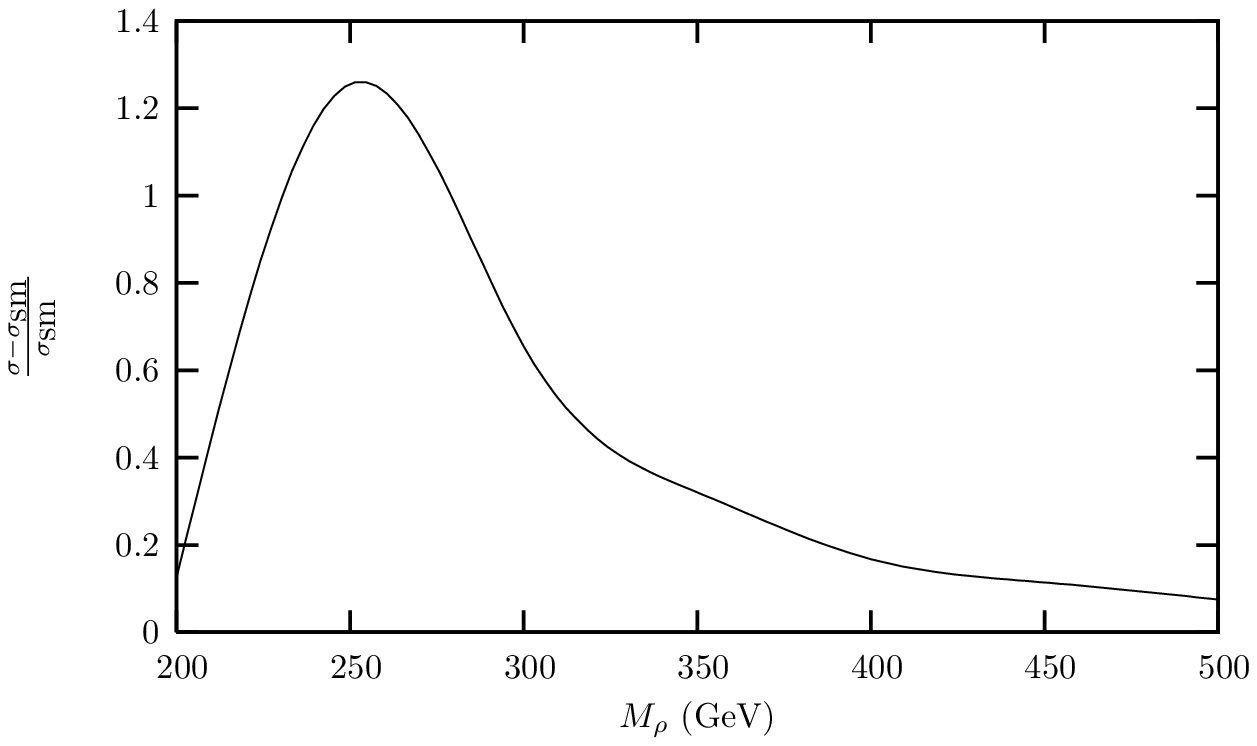}
  \caption{Enhancement of the cross section in the process $p\bar{p}
  \rightarrow W^+ H$ at the Tevatron for $M_H=150$ GeV }
  \label{fig:tev}
\end{figure}

We also compute the cross section of the process $p\bar{p} \rightarrow
H W^+$ at the Tevatron (Run II). The result is shown in figure
\ref{fig:tev} for $M_H=150$ GeV. Notice that the enhancement in this
case is comparable to the prediction for the LHC.

The point in the parameter space we use for studying our model was
chosen in order to maximize the deviation from the Standard Model for
the selected channel. This procedure allows us to evaluate the
possibility of testing the model. Let's now consider the restrictions
imposed by data on the electroweak parameters $S, T$ and $U$. This is
not the place for performing a detailed calculation of these
parameters in our model, nevertheless we expect that the final result
must be similar to the one obtained in the minimal BESS model. 
\cite{dominici,mbess}. Written in the notation of the original
authors, this result is:
\begin{eqnarray*}
  \epsilon_1&=&0\\
  \epsilon_2&=&0\\
  \epsilon_3&=&-\frac{b}{2}+\left(\frac{g}{g''}\right)^2
\end{eqnarray*}
   where $\epsilon_1,\epsilon_2,\epsilon_2$ are proportional to $T,U$
   and $S$ respectively, $g''$ is our coupling constant $g_2$ and $b$
   corresponds to our parameter $x_1$ which represents a direct
   coupling of the proto-technirho to fermions. For the set of
   parameters used above   we obtain 
   $\epsilon_3=0.01$ which is disfavored by precision
   measurements. Nevertheless, we can be consistent with the constrains
   imposed by precision data by choosing $x_1=2(g/g_2)^2$. In this case,
   our results on $\sigma-\sigma_{\mbox{\tiny{SM}}}$ are modified by a
   factor $0.60$ and an important enhancement remains in the channel
   $pp \rightarrow H W^+$  for $M_{\rho}$
   around $250$ GeV. 

At this point a word must be said about the direct search of our
vector resonances and the mass limits imposed to them by current
data. In general the results of direct search, performed at the
Tevatron and LEP, of the technirho and techniomega predicted in usual walking
technicolor models applies to our case, except by the fact that we have
suposed there are no physical technipions in the spectrum. In this
case, considering the technirho decay to charged
leptons and a pair of $W$'s, the technirho (as well as the
techniomega) is excluded for $M_{\rho}<206$ GeV\cite{pdg}.        

\section{Conclusions}
\label{sec:con}

We have constructed an effective lagrangian which represents the
Standard Model with a light higgs bosons and vector resonances that
mix with the gauge bosons. We fixed the parameter of the model that
connects the mass of the new vector bosons with their coupling
constant, in such a way that the model were compatible with light
resonances.

We want to emphasize that, although this effective lagrangian is
inspired by a new technicolor scenario recently proposed in
\cite{newmodels}, we do not claim that the values chosen for our study
 represent the low energy limit of the specific models constructed there. 

 The most obvious process for searching differences between
our model and the predictions of the Standard Model is the associate
production of a higgs and a gauge boson. We found that the most
sensitive channel is the production of the higgs and $W$.

For a range of resonance's mass between $200$ GeV and $350$ GeV the
enhancement of the cross section is significant at both, the LHC and
the Tevatron.

\section*{Acknowledgements}

The author is grateful to Rog\'erio Rosenfeld for valuable comments

\newpage

\begin{thebibliography}{99}

\bibitem{technicolor}
For a recent review see 
C.~T.~Hill and E.~H.~Simmons,
  Phys.\ Rept.\  {\bf 381}, 235 (2003)
  [Erratum-ibid.\  {\bf 390}, 553 (2004)]
  [arXiv:hep-ph/0203079].
\bibitem{newmodels}
D.~D.~Dietrich, F.~Sannino and K.~Tuominen,
  Phys.\ Rev.\ D {\bf 72}, 055001 (2005)
  [arXiv:hep-ph/0505059].\\

\bibitem{dominici}
For a review of effective models of a strong electroweak symmetry
breaking sector with scalar and vector resonances, see
D.~Dominici,
  Riv.\ Nuovo Cim.\  {\bf 20}, 1 (1997)
  [arXiv:hep-ph/9711385].

\bibitem{VMD}
For a modern review on Vector Meson Dominance see,
H.~B.~O'Connell, B.~C.~Pearce, A.~W.~Thomas and A.~G.~Williams,
Prog.\ Part.\ Nucl.\ Phys.\  {\bf 39}, 201 (1997)
[arXiv:hep-ph/9501251].

\bibitem{zerwekh_and_rosenfeld}
A.~R.~Zerwekh and R.~Rosenfeld,
Phys.\ Lett.\ B {\bf 503}, 325 (2001)
[arXiv:hep-ph/0103159].

\bibitem{zerwekh}
A.~R.~Zerwekh,
  arXiv:hep-ph/0307130.

\bibitem{Chivukula:2005bn}
  R.~S.~Chivukula, E.~H.~Simmons, H.~J.~He, M.~Kurachi and M.~Tanabashi,
  Phys.\ Rev.\ D {\bf 71}, 115001 (2005)
  [arXiv:hep-ph/0502162]

\bibitem{bando}M. Bando, T. Kugo e K. Yamawaki, \PRT{164}{88}{217}.

\bibitem{mbess}R. Casalbuoni, S. De Curtis, D. Dominici and  R. Gatto,
              \PLB{155}{85}{95}; \NPB{282}{87}{235};\\
              R. Casalbuoni, P. Chiappetta, A. Deandrea, D. Dominici
              and  R. Gatto, \ZPC{60}{93}{315};\\
              R. Casalbuoni, P. Chiappetta, S. De Curtis, F. Feruglio,
              R. Gatto,  B. Mele and J. Terron, \PLB{249}{90}{130};\\
              R. Casalbuoni, P. Chiappetta,M.C. Cousinou, S. De
              Curtis, F. Feruglio, R. Gatto, \PLB{253}{91}{275};\\
              L. Antichini,R. Casalbuoni and S. De Curtis,\PLB{348}{95}{521}.


\bibitem{lbess}
  R.~Casalbuoni, S.~De Curtis, D.~Dominici and M.~Grazzini,
  Phys.\ Rev.\ D {\bf 56}, 5731 (1997)
  [arXiv:hep-ph/9704229].

\bibitem{LanHEP}
A.~V.~Semenov,
  arXiv:hep-ph/0208011.
\\
A. Semenov. LanHEP - a package for automatic generation of Feynman
rules. User's manual. INP MSU Preprint 96-24/431, Moscow, 1996;
hep-ph/9608488 \\ 
A. Semenov. Nucl.Inst.\& Meth. A393 (1997) p. 293. \\
A. Semenov. LanHEP - a package for automatic generation of Feynman
rules from the Lagrangian. Updated version 1.3. INP MSU Preprint
98-2/503. \\
Home page:http://theory.sinp.msu.ru/~semenov/lanhep.html

\bibitem{CompHEP}
E.Boos et al. [CompHEP Collaboration], CompHEP 4.4: Automatic
computations from Lagrangians to events,
Nucl. Instrum. Meth. A534(2004), p250 [hep-ph/0403113].\\ 
 CompHEP - a package for evaluation of Feynman diagrams and integration
over multi-particle phase space. User's manual for version 3.3,
hep-ph/9908288 \\
 Home page: http://theory.sinp.msu.ru/comphep


\bibitem{rosenfeld}R.~Rosenfeld,
  Phys.\ Rev.\ D {\bf 50}, 4283 (1994)
  [arXiv:hep-ph/9403356].


\bibitem{pdg}S. Eidelman et al, Phys. Lett. B 592, 1 (2004).
\end{thebibliography}
\end{document}